\documentclass[superscriptaddress,aps,prl,floats, twocolumn,twoside,floatfix]{revtex4}
\usepackage{epsfig}
\usepackage{amsmath}
\usepackage{amssymb}
\usepackage[english]{babel}

\newcommand{\be}{\begin{equation}}
\newcommand{\ee}{\end{equation}}
\newcommand{\bea}{\begin{eqnarray}}
\newcommand{\eea}{\end{eqnarray}}

\begin{document}

\title{On the structure of correlations in the three dimensional spin glasses.}  

\author{Pierluigi Contucci}
\affiliation{
Universit\`{a} di Bologna, Piazza di Porta S.Donato 5, 40127 Bologna, Italy}

\author{Cristian Giardin\`a}
\affiliation{Technische Universiteit Eindhoven and EURANDOM, P.O.Box
513, 5600MB Eindhoven, Netherlands}

\author{Claudio Giberti}
\affiliation{ Universit\`a di Modena e Reggio Emilia, via G.
Amendola 2 -Pad. Morselli- 42100 Reggio Emilia, Italy}

\author{Giorgio Parisi}
\affiliation{
Universit\`a La Sapienza di Roma,  Roma, Italy, CNR-INFM SMC and INFN, sezione di Roma.}

\author{Cecilia Vernia}
\affiliation{
Universit\`{a} di Modena e Reggio Emilia, via Campi 213/B, 41100 Modena, Italy}

\begin{abstract}
We investigate the low temperature phase of
three-dimensional Edwards-Anderson model with Bernoulli random couplings.
We show that at a fixed value $Q$ of the overlap the model fulfills the clustering
property: the connected correlation functions between two local overlaps decay
as a power whose exponent is independent of $Q$ for all $0\le |Q| < q_{EA}$.
Our findings are in agreement with the RSB theory and show that the
overlap is a good order parameter.
\end{abstract}

\maketitle

Spin glasses have unusual statistical  properties. In mean field
theory there are intensive quantities that fluctuate also in the
thermodynamic limit. This is the effect of the coexistence of
many equilibrium states. The correlations functions inside a given
state should have a power law behavior: below the critical
temperature spin glasses are always in a critical state (many
glassy systems should share this behavior). These predictions of
mean field theory have been never studied in details, apart from
\cite{DGMMZ}; the aim of this paper is to address this point in a
careful way.

In order to better characterize the behavior of spin glasses, it is
convenient to consider two replicas, or clones, of the same system
(let us call them  $\sigma(i)$ and $\tau(i)$ where $i$ denotes the
point of the lattice). The two clones share the same Hamiltonian
$H_J$, the label $J$ indicates the set of random coupling constants.

Let us define the local overlap $ q(i)\equiv \sigma(i) \tau(i) $ and
the global overlap $q\equiv V^{-1}\sum_{i} q(i)$, $V$ being the
total volume. For the three dimensional EA model \cite{EA} at zero magnetic
field (to be defined later) all simulations confirm that the
probability distribution of $P_J(q)$ is non trivial in the
thermodynamic limit, it changes from system to system and its
average over the disorder, that we denote as $P(q)\equiv E[
P_J(q)]$, is non trivial and it has a support in the region from
$-q_{EA}$ to $q_{EA}$, $q_{EA}$ being the overlap of two generic
configurations belonging to the same state. It is usually
assumed that the function $P(q)$ has in the infinite volume limit a
delta function singularity at $q=q_{EA}$, that appears as a peak in
finite volume systems. In the presence of multiple states the most
straightforward approach consists in identifying the clustering
states (i.e. those where the connected correlation functions go to
zero at large distance) and to introduce an order para\-meter that
identifies the different states. This task is extremely difficult in
a random system where the structure of the states depends on the
instance of the system. However the replica theory is able to make
predictions  without  finding out explicitly  the set of states. At
this end the introduction of the two clones plays a crucial role.
Indeed if the global overlap $q$ has a preassigned value, the theory
predicts that the correlations of local overlaps $q(i)$ go to zero
at large distances. In other words $q$ is a good order parameter.

For each realization of the system we consider two clones. The
observables are the local overlaps $q(i)$ and their correlations. We
define $\langle O \rangle^J_Q$ the expectation value of the
observable $O$ in the $J$-dependent Gibbs ensemble restricted to
those configurations of the  two clones that have global overlap
$q=Q$. We define the average  expectations values $\langle O
\rangle_Q$ as weighted average over the systems of restricted
expectation values \footnote{Alternatively we could define  the
expectations values $\langle O \rangle_Q$ as the unweighted average
over the systems of restricted expectation values: $ \langle O
\rangle_Q=E[\langle O \rangle^J_Q] \ .\label{KO} $ We followed
definition in eq. (\ref{OK}), that is very easy to implement
numerically: all the configurations produced in a numerical
simulation are classified according to their overlaps independently
from the system they come from and we perform the average of each
class.}: {\bf
\begin{equation}
\langle O \rangle_Q={E[P_J(Q) \langle O \rangle^J_Q] \over E[P_J(Q)]}\ . \label{OK}
\end{equation}
}

The main statement of the RSB theory \cite{MPV} (to which we bring evidence in
this paper) is that the $Q$-dependent connected correlation
functions go to zero when computed in the ensemble $\langle \cdot
\rangle_Q$, i.e. the states $\langle \cdot  \rangle_Q$  are
clustering. The procedure is very similar to the one used in
ferromagnetic models to construct clustering states (for example in
the Ising case by considering averages with positive, or negative,
total magnetization). The overlap constraint state is not clustering when the
equilibrium state is locally unique (apart from a global change of
signs) \cite{FH}. Indeed this is the only known example of a system where the
clustering states are labeled by a continuously changing order
parameter in absence of a continuous symmetry, like rotations or
translations.

This clustering property has far reaching consequences: for
example the probability distribution $P(q)^W_Q$ of the window
overlaps \cite{CINQUE}, i.e. the average overlap over a region of
size $W$, becomes a delta function in the infinite volume limit: $
\lim_{W \to \infty}  P(q)^W_Q=\delta(Q-q).$ Indeed when the region
$W$ becomes large, the window overlaps are intensive quantities that
do not fluctuate inside clustering states.

It is convenient to  recall the known theoretical results for the
connected correlation functions in the case of short range Ising
spin glasses:
\begin{equation}
G(x|Q)=\langle q(x) q(0)\rangle_Q,\ \ \ \ \ \ C(x|Q)=G(x|Q)- Q^2
\end{equation}
and their Fourier transforms $\tilde C(k|Q)$.
\begin{figure}
 \includegraphics[width=9cm,height=6cm]{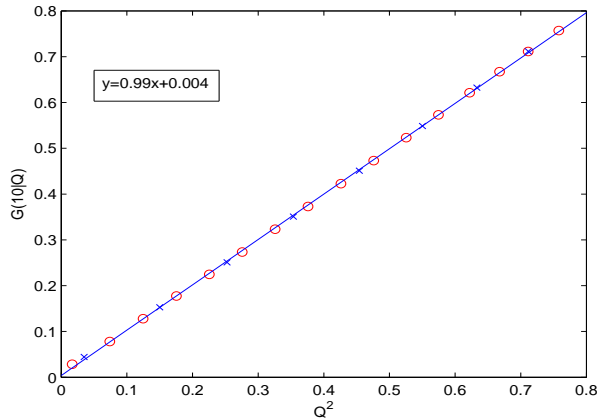}
 \caption{The correlation function at distance 10 $G(10|Q)$ averaged
 in 20 bins (round points) and in ten bins (crosses) of $Q^2$, versus
 the average of $Q^2$ inside the bin. The data are for a system of size
 20 and the straight is the best fit to the data. } \label{LINEAR}
\end{figure}

The simplest  predictions are obtained starting from mean field
theory and computing the first non trivial term \cite{DKT}
Neglecting logarithms we have in the small $k$ region
\begin{eqnarray}
\tilde C(k|Q) \propto k^{-4} \ \ \  &\mbox{for}& \ \ \ \ Q=0 \ , \nonumber \\
\tilde C(k|Q) \propto k^{-3} \ \ \  &\mbox{for}& \ \ \ \ 0<Q<q_{EA} \ , \nonumber \\
\tilde C(k|Q) \propto k^{-2} \ \ \  &\mbox{for}& \ \ \ \ Q=q_{EA}  \ , \nonumber \\
\tilde C(k|Q) \propto (k^{2}+\xi(Q)^{-2} )^{-1} \ \ \ &\mbox{for}& \ \ \ \ Q>q_{EA}  \ .
\end{eqnarray}
These results are supposed to be exact at large distances in
sufficiently high dimensions, i.e. for $D>6$.

The reader may be surprised to find a result for $Q>q_{EA}$ because
the function $P(q)$ is zero in this region in the infinite volume
limit. However for finite systems $P(q)$ is different from zero for
any $q$ albeit it is very small \cite{FPV,BFM} in the region
$q>>q_{EA}$. For $Q>q_{EA}$ an analytic computation of the function
$\tilde C(k|Q)$ has not yet been done, however it is reasonable that
the leading singularity near to $k=0$ in the complex plane is a
single pole, leading to an exponentially decaying correlation
function \footnote{An exponential decrease of the correlation is
present in the Heisenberg model if we constraint the modulus of
magnetization to have a value greater than the equilibrium value. On
the contrary if the modulus of the magnetization is less than the
equilibrium value, the connected correlation function does not go to
zero at large distances.}.

When the dimensions become smaller than $6$ we can rely on the
perturbative expansion in $\epsilon=6-D$, where only the first order
is (partially) known \cite{TD}. It seems that  predictions at
$Q=q_{EA}$ should not change and the form of the $k=0$ singularity
at $Q=q_{EA}$ (i.e. when the two clones belong to the same state)
remains $k^{-2}$ as for Goldstone Bosons. On the contrary the $k=0$
singularities at $Q<q_{EA}$ should change. We can expect that
\begin{equation}
\tilde C(k|Q) \propto k^{-\tilde\alpha(Q)} \ \ \  \mbox{for} \ \ \ \
0\le Q<q_{EA}\ .
\end{equation}
These perturbative results are the only information we have on the
form of $\tilde\alpha(Q)$. The simplest scenario would be that $\tilde\alpha(Q)$
is discontinuous at $Q=0$ and constant in the region $0<Q<q_{EA}$.
It would be fair to say that there is no strong theoretical evidence
for the constancy of $\tilde\alpha(Q)$ in the region $0<Q<q_{EA}$,
apart from generic universality arguments. On the contrary the
discontinuity at $Q=0$ could persist in dimensions not too smaller
than 6, and disappear at lower dimensions, as supported by our data
in $D=3$.

In three dimensional case in configuration space we should have:
\begin{equation}\label{POWER}
C(x|Q)\propto x^{-\alpha(Q)}  \ \ \  \mbox{for} \ \ \ \ 0\le Q\le
q_{EA}\ .
\end{equation}
with $\alpha(q_{EA})=1$ (indeed in general we have that $\alpha(Q) = D -
\tilde\alpha(Q)$). For $Q>q_{EA}$ the correlation should go to zero
faster than a power: we tentatively assume that
\begin{equation}
C(x|Q)\propto x^{-1} \exp(-x/\xi(Q)) \ \ \  \mbox{for} \ \ \ \  q_{EA} <Q\ ,
\end{equation}
(also other behaviours are possible).

In this paper we will numerically study the properties of the two
overlaps connected correlations functions in the three dimensional
EA model. The Hamiltonian of the EA model \cite{EA} is given by
$
H_\sigma=-\sum_{|i-j|=1}J_{i,j}\sigma_i\sigma_j
$
with $J_{i,j}=\pm 1$ (symmetrically distributed) and Ising spins $\sigma_i = \pm 1$.

\begin{figure}
 \includegraphics[width=9cm,height=6cm]{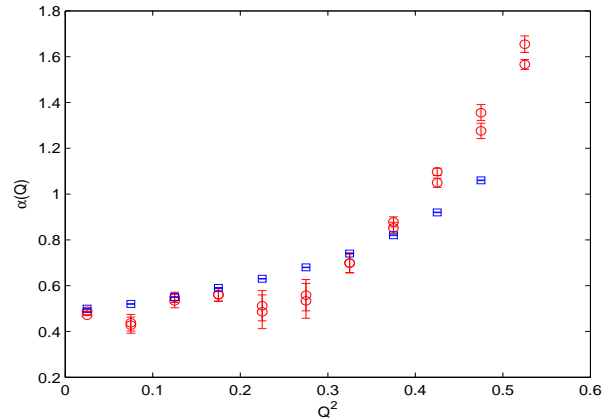}
 \caption{Circles are the value of the exponent $\alpha(Q)$ by
 fitting $\chi^{(s)}(Q)_L$ as a power of $L$: for each value of $Q^2$
 we show two points corresponding to $s=1$ and to $s=2$. Squares
 represent the same quantity $\alpha(Q)$ as obtained through the
 scaling approach visible in fig. \ref{COLLAPSE}.} \label{EXP}
\end{figure}

We have studied cubic lattice systems with periodic boundary
conditions of side $L$ for $L=4,6,8,10,12,16,20$. The simulation
parameters are the same used in reference \cite{CGGPV}. We present
the results only at temperature $T=0.7$, while the critical
temperature is  about $T=1.11$.

We have first classified the configurations created during the
numerical simulations according to the value of the global overlap
$q$. Since the properties of the configurations are invariant under
the symmetry ($q \to -q$), we have classified the configurations
into 20 equidistant bins in $q^2$: e.g. the first bin contains all
the configurations where $0<q^2<1/20$. In this way we compute the
correlations $C(x|Q)$. We have measured the correlations only along
the axes of the lattice: $x$ is an integer restricted to the range
$0,L/2$. As a control we have done the same operation with 10 bins
obtaining similar results.

We have firstly verified that the connected correlations vanish for
large systems. At this end in fig. (\ref{LINEAR}) we have plotted
for $L=20$ (our largest system) the correlation $G(10|Q)$ versus the
average of $Q^2$ in the bin. We see that the two quantities
coincide. The data show a strong evidence for the vanishing of the
connected two point correlation function. The predictions of replica
theory is $G(10|Q)=Q^2$, neglecting small corrections going to zero
with the volume.

Further information can be extracted from the data. The analysis of
the data should be done in a different way in the two regions $0\le
Q^2\le q_{EA}^2$ and $ q_{EA}^2 <Q^2$ as far as two different
behaviour are expected. In our case $q_{EA}^2$ can be estimated to
be around $0.4$.

\begin{figure}
 \includegraphics[width=9cm,height=6cm]{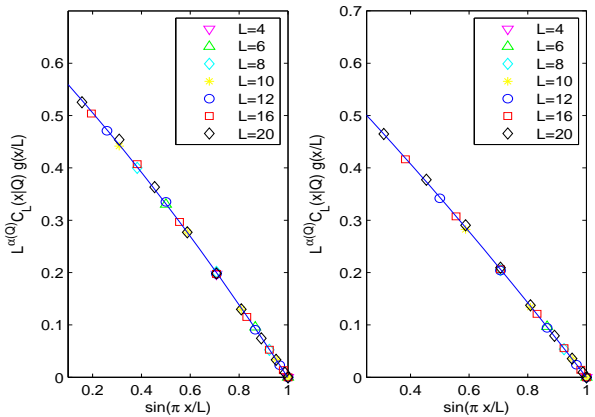}
 \caption{ The quantity $L^{\alpha(Q)}C_L(x|Q) g(x/L)$ with
 $g(z)=(1/z+1/(1-z))^{-\alpha(Q)}$ versus $\sin(\pi x/L)$
 for the set of data in the first bin $Q^2<0.5$, using the best
 value of $\alpha(Q)$. The left panel display the data for all
 the correlation at distances $x\ge1$ (the corresponding value
 of $\alpha(Q)$ being 0.50); in the right panel we have only the data
 with $x\ge2$  (the corresponding value of $\alpha(Q)$
 being 0.54). } \label{COLLAPSE}
\end{figure}

In the region $0\le Q^2\le q_{EA}^2$ the power law decrease (\ref{POWER})
of the correlation is expected. To test this hypothesis \cite{JANUS} we
define for each $L$ the quantities $\chi^{(s)}_L(Q)$
\begin{equation}
\chi^{(s)}_L(Q)=\sum_{x=1}^{L/2} x^s C_L(x|Q).
\end{equation}
where  $C_L(x|Q)$ is the connected correlation function in a system
of size $L$, i.e. $G_L(x|Q)-Q^2$ \footnote{In order to decrease the
statistical errors we have used the asymptotically equivalent
definition of the connected correlation function
$G_L(x|Q)=G_L(x|Q)-G_L(L/2|Q)$.}. For large $L$  $\chi^{(s)}_L(Q)$
should behave as $L^{s+1-\alpha(Q)}$. We have evaluated the previous
quantity \footnote{More precisely we have used at the place of
$C_L(x|Q)$ its proxy $C_L(x|Q)-C_L(L/2|Q)$, that has smaller
statistical errors.} for $s=1,2$. In the region $Q^2<0.4$ we have
found that the ratio $\chi^{(2)}_L(Q)/\chi^{(1)}_L(Q)$ is well
linear in $L$. Here the data for  $\chi_L^{(s)}(Q)$ can be well
fitted as a  power of $L$ and the exponents $\alpha(Q)$
computed using $s=1$ and $s=2$  coincide within their errors. These
results are no more true in the region $0.5<Q^2$ indicating that a
power law decrease of the correlation is not valid there. The
exponents we find with this method are shown in fig. (\ref{EXP}).

In order to check these results for $\alpha(Q)$ we have used a
different approach. In the large volume limit the correlation
function should satisfy the scaling
\begin{equation}
L^{\alpha(Q)}C_L(x|Q)= f(x/L). \label{SCALE}
\end{equation}
The value of $\alpha(Q)$ can be found by imposing this scaling. At
this end for each value of $Q$ we have plotted
$L^{\alpha(Q)}C_L(x|Q)$ and found the value of $\alpha(Q)$ for which
we get the best collapse. The result of the collapse is shown in
fig. (\ref{COLLAPSE}) for $Q$ around zero, where for graphical
purpose we have plotted $L^{\alpha(Q)}C_L(x|Q) g(x/L)$ versus
$\sin(\pi x/L)$, where the function $g$ has been added to compress
the vertical scale (we find convenient to use
$g(z)=(1/z+1/(1-z))^{-\alpha(Q)}$, following \cite{MaPa}). In the
left panel we show the collapse using all points with $x\ge 1$, in
the right panel we exclude the correlations at distance $x=1$. The
corresponding values of the exponent are shown in fig. (\ref{EXP})
and they agree with the ones coming from the previous analysis in
the region of $Q^2\le 0.4$.

The exponent $\alpha(Q)$ is a smooth function of $Q^2$ which goes to
1 near $Q^2=0.4$ in very good agreement with the theoretical
expectations. We do not see any sign of a discontinuity at $Q=0$,
and this is confirmed by an analysis with an high number of bins
(e.g. 100). However it is clear that for lattice of this size value
we cannot expect to have a very high resolution on $Q$ and we should
look to much larger lattices in order to see if there is a sign of a
building up of a discontinuity and of a plateaux.  The value of the
exponent that we find at $Q=0$ is consistent  with the value 0.4
found from the dynamics \cite{JANUS}, and with the value 0.4 found
with the analysis of the ground states with different boundary
conditions \cite{MaPa}.
\begin{figure}

\includegraphics[width=1.0\columnwidth]{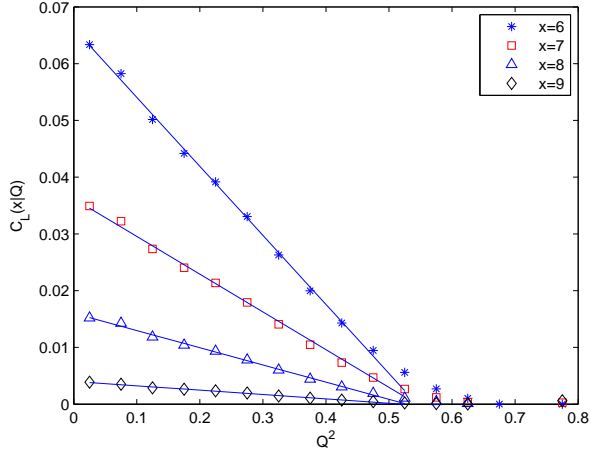}
\caption{The connected correlation $C_L(x|Q)$ as function of $Q^2$
for $x=6,7,8,9$ at $L=20$. The straight lines are linear fits.}
\label{CORRELIN}
\end{figure}

From the previous analysis it is not clear if the exponent
$\alpha(Q)$ has a weak dependence on $Q$ or if the weak dependence
on $Q$ is just a pre-asymptotic effect. In order to clarify the
situation it is better to look to the connected correlations
themselves. In fig. (\ref{CORRELIN}) we display the connected
correlation $C_L(x|Q)$ as function of $Q^2$  for $x=6,7,8,9$ at
$L=20$, our largest lattice (for the result at $x=1$ see \cite{CGGV}).
We can fit the correlations at fixed
$L$ (e.g. $L=20$) for large $x$ as:
\begin{equation}
C_L(x|Q)= A(x,L)(Q^{2}-B(x,L)^{2}) \, , \ \ Q^{2}<B(x,L)^{2}
\label{WOH}
\end{equation}
while $C_L(x|Q)$ is very near to zero for  $ Q^{2}>B(x,L)^{2}$. The
goodness of these fits improves with the distance (similar results
are valid at smaller $L$). The value of $B(x,L)^{2}$ is near to
$q_{EA}^2$ and it is slightly decreasing with $L$. The validity of
the fits (\ref{WOH}) for large $L$ would imply that in the region
$|q|<q_{EA}$ the large distance decrease of the correlation function
should be of the form $A(Q^{2}-q_{EA}^{2})f(x)$ and therefore the
exponent $\alpha(Q)$ should not depend on $Q$.

\begin{figure}
 \includegraphics[width=1.0\columnwidth]{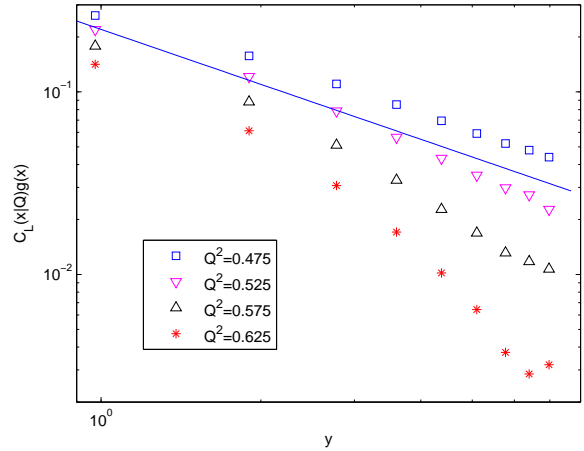}
 \caption{Correlations functions at $L=20$ for
 $q^{2}$ = 0.475, 0.525, 0.575, 0.625, versus
 $y\equiv(1/x+1/(2L-x))^{-1}$. The straight line is
 proportional to $y^{-1}$. } \label{CROSS}
\end{figure}

However near $q=q_{EA}$ we should have a real crossover region.
In Fig. \ref{CROSS} we
show $C_L(x|Q)g(x)$ at $L=20$ for $0.475\le Q^{2}\le 0.625$, versus
$y\equiv(1/x+1/(2L-x))^{-1}$ (we use the variable $y=x(1-O(x/L))$ to
take care of finite volume effects) with $g(x)=(1-2x/L)^{-2}$.
It seems that the data at $Q^{2}>0.475$ decrease faster than a power at
large distances and that the data at $Q^{2}=0.475$ are compatible with
a power with exponent $-1$ It is difficult to extract precise quantitative
conclusions, without a careful analysis of the $L$ dependence. We
hope that this will be done when the data on the correlation
functions on larger lattices will be available.

In the region $q_{EA}^2\le Q^2$ our task is different:  the
correlations are short range and we would like to compute if
possible the correlation length. At this end we have fitted the data
as
\begin{equation}
C_L(x|Q)={a \over x+1} \exp (-x/\xi_L(Q))+ (x \to L-x) +const
\label{DECAY}
\end{equation}
The choice of the fit is somewhat arbitrary, however we use it only
to check that the correlation length diverges at $q_{EA}$ and that
near $q_{EA}$ are well fitted by a $1/x$ power. The fits are good,
but this may not imply the correctness of  the functional form in
eq. (\ref{DECAY}). We find that far from $Q=0.5$ the correlation
length is independent of $L$, (it is quite small). We have tried to
collapse the data for $L>8$ in the form
\begin{equation}
\xi_L(Q)=L\, f((Q^2- q_{EA}^2)L^{1/\nu})\ .
\end{equation}
A reasonable collapse has been obtained, however the $q_{EA}^2$ is
quite small (i.e. 0.25): it is quite possible that there are finite
volume effects and different ways to evaluate $q_{EA}$ give
different results on a finite lattice and they would converge to the
same value in the infinite volume limit.

In conclusion the global overlap for a two clone system is a well
defined order parameter such that in the appropriate restricted
ensemble the two points connected correlation function decays at
large distance. The connected correlations decay as a power whose
exponent seem to be independent from $Q$ for $0\le|Q|<q_{EA}$: the value
of the exponent is in agreement with the results obtained in
different context at $Q=0$. Moreover the connected two points
correlation functions at $Q=q_{EA}$ decays like $1/x$ in agreement
with the detailed predictions coming from replica theory.

{\bf Acknowledgments.} We thank E.Marinari.
P. Contucci acknowledge STRATEGIC RESEARCH GRANT from University
of Bologna.
C. Giardin\`a and C. Vernia acknowledge GNFM-INdAM for partial
financial support.

\end{document}